\documentclass[aps,prb,superscriptaddress,twocolumn,floatfix,showpacs,citeautoscript,longbibliography]{revtex4-1}

\usepackage{graphicx}
\usepackage{dcolumn}
\usepackage[tight]{subfigure}
\usepackage{amsmath}
\usepackage{amssymb}
\usepackage{verbatim}
\usepackage{color}
\usepackage{bm} 
\usepackage{natbib}
\usepackage{xspace}
\usepackage{mathtools}
\usepackage{physics}
\usepackage{hyperref} 
\hypersetup{colorlinks=true,linktoc=all,linkcolor=blue,breaklinks=true,citecolor=blue,urlcolor=blue}
\usepackage{etoolbox}
\usepackage{epstopdf}

\begin{document}
\title{Dynamics of the Anderson impurity model: benchmarking a non-adiabatic exchange-correlation potential in time-dependent density functional theory}

\author{Niklas Dittmann}
\affiliation{Institute for Theory of Statistical Physics, RWTH Aachen, 52056 Aachen, Germany and JARA -- Fundamentals of Future Information Technology}
\affiliation{Department of Microtechnology and Nanoscience (MC2), Chalmers University of Technology, SE-41298 G{\"o}teborg, Sweden}
\author{Nicole Helbig}
\affiliation{Institute for Theory of Statistical Physics, RWTH Aachen, 52056 Aachen, Germany and JARA -- Fundamentals of Future Information Technology}
\author{Dante M. Kennes}
\affiliation{Dahlem Center for Complex Quantum Systems and Fachbereich Physik, Freie Universit{\"a}t Berlin, 14195 Berlin, Germany}

\pacs{}

\begin{abstract}
In this comparative study we benchmark a recently developed non-adiabatic exchange-correlation potential within time-dependent density functional theory (TDDFT) (Phys.\ Rev.\ Lett.\ {\bf 120}, 157701 (2018)) by (a) validating the transient dynamics using a numerically exact density matrix renormalization group approach as well as by (b) comparing the $RC$-time, a typical linear response quantity, to upto second order perturbation theory results. As a testbed we use the dynamics of the single impurity Anderson model. These benchmarks show that the non-adiabatic functional yields quantitatively accurate results for the transient dynamics for temperatures of the order of the hybridization strength while the TDDFT $RC$-times quantitatively agree with those from second-order perturbation theory for temperatures which are large compared to the hybridization strength. Both results are particularly intriguing given the relatively low numerical cost of a TDDFT calculation (at least compared to exact approaches).
\end{abstract}
\maketitle
\section{Introduction}

Nowadays, non-equilibrium quantum many-body physics is at the vanguard of contemporary condensed matter physics. The interest in non-equilibrium phenomena has been stimulated by the ability to conduct controlled experiments, in which external parameters can be influenced with great precision. Several of the first realizations of setups with unprecedented control opportunities were using quantum dots to confine the electrons \cite{Hanson07}. Describing quantum dot experiments out of equilibrium from a theoretical side has hence become a vital field of research. The simplest description of a quantum dot harboring a single energy level employs the so-called Anderson impurity model,\cite{Anderson61} in which a single correlated spinful degree of freedom is coupled to a Fermi-liquid reservoir of particles, see Fig.\ \ref{fig_setup}(a). The Anderson impurity model out-of equilibrium is the subject of a plethora of theoretical investigations using different tools such as perturbative approaches\cite{Schoeller96}, numerical\cite{Costi17} or density matrix renormalization group calculations\cite{Wolf14}, machine learning ans\"atze\cite{Arsenault14}, quantum Monte-Carlo simulations\cite{Gull11} or time-dependent density functional theory\cite{Uimonen11} to name a few. Each of these approaches has their merits as well as shortcomings: some of them are numerically exact, but are difficult to apply to multi-quantum dot geometries or more complicated couplings (such as spin-orbit coupling), while others employ approximations but tend to generalize more easily. 

In order to develop accurate methods for treating more elaborate quantum-dot setups, the approximate approaches require a thorough benchmarking with exact results when possible. This is an important step to establish the range of validity and applicability of the different (and often complementary) approaches.\cite{Eckel10} In Ref.~\onlinecite{Uimonen11} the authors conduct such a comparison between time-dependent density functional theory (TDDFT) and the density matrix renormalization group (DMRG) (and perturbative methods). The Hartree-exchange-correlation (HXC) potential entering the TDDFT approach was approximated using an adiabatic local density approximation, which was motivated from Bethe-ansatz insights obtained for the Hubbard model. It was shown that this approximation accurately describes the electron density on the quantum dot but steady-state currents are overestimated. The authors conjectured that improvements require the functional to be nonlocal in space and time \cite{Uimonen11,Kurth10}.

To improve the performance of TDDFT, recently, a non-adiabatic approximation for the Hartree-exchange-correlation potential of the single-impurity Anderson model was derived by exploiting analogies to quantum transport theory \cite{Dittmann2018}. The derivation is based on a first-order perturbative treatment in the tunnel coupling between the impurity and the reservoir and uses a Markov approximation for the time propagation in the rate equation approach. It was shown that the resulting non-adiabatic functional improves over its adiabatic counterpart and yields the correct exponential decay of the density after a quench in the gate voltage, see Fig.\ \ref{fig_setup}(b). Directly after the quench, the decay of the density in the TDDFT description deviates from the one obtained with the rate equation. This difference was attributed to the Markov approximation being made in the rate equation while the TDDFT time propagation did not suffer from this additional approximation. However, in TDDFT one propagates the Kohn-Sham system with an approximate HXC potential. Hence, it is not clear which of the two methods describes the short to intermediate time behaviour more accurately. In the present work, we compare the TDDFT description of this transient dynamics to time-dependent DMRG results. While DMRG provides numerically exact results, it is a low-entanglement method scaling exponentially in the amount of entanglement in the system under scrutiny. In typical quench scenarios, entanglement build-up is linear in simulation time and, as a consequence, numerical resources are exhausted in an exponential fashion. Depending on the prefactor of entanglement growth this restricts accessible time-scales to short-to-intermediate ones. For more complex quantum dot geometries the overall scaling of numerical cost becomes even worse. In summary, DMRG is the perfect tool for benchmarking the short to intermediate time behaviour of the non-adiabatic approximation in TDDFT such that we can then address more complex systems within TDDFT with confidence. As a second test, we compare the $RC$-times calculated from linear-response TDDFT with those from first and second-order perturbative treatments in the tunnel coupling. While the approximate exchange-correlation potential in TDDFT was derived using only the first order in the tunnel coupling, higher orders can enter due to the exact time propagation of the Kohn-Sham system. 

Finally, we note that pushing the boundaries of methods available to describe the dynamics of the Anderson impurity model is relevant also beyond the description of quantum dot dynamics in experiments. Solutions or high quality approximations to the quantum impurity problem out-of-equilibrium are also urgently needed as impurity solvers in dynamical mean field theories (DMFT) and its variances \cite{Gull11}, which are nowadays at the frontier of strongly correlated condensed matter research. Especially treating more complicated quantum dot-geometries and spin-orbit coupling is a subject of increasing recent research attention. Furthermore, obtaining a deeper understanding of non-equilibrium physics in general and specifically the transient response to external changes of quantum dots and beyond is also crucial to efficiently harvest the promises made by the blossoming field of quantum technologies.\cite{Zwanenburg13} 

The paper is structured as follows: In section \ref{sec:methods} we introduce the Anderson impurity model that is used for all the calculations and present the two methods, time-dependent density functional theory and density matrix renormalization group theory. We discuss the results for the transient dynamics that were obtained with these methods in section \ref{sec:results} where we also present the comparison of the $RC$ times from TDDFT linear response with those from first and second order perturbation theory. We conclude our work in section \ref{sec:conclusions}.

\section{Model and Methods}\label{sec:methods}

\subsection{Model}\label{sec:model}
\begin{figure}[t]
\begin{center}
\includegraphics[width=\columnwidth]{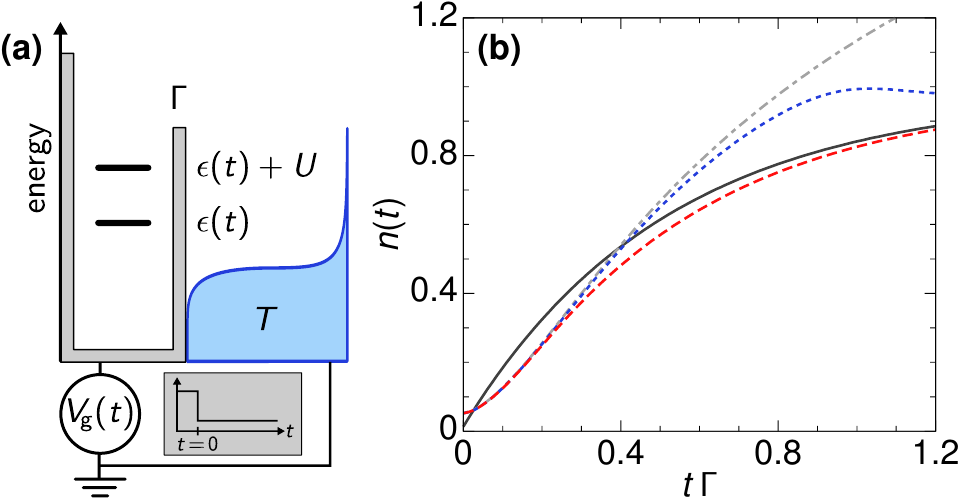}
\end{center}
	\caption{	
 	(a) Sketch of the single impurity Anderson model at finite temperature $T$.
 	The single energy-level $\epsilon(t)$ with on-site interaction $U$ is tunnel coupled to an electron reservoir with tunnel-coupling strength $\Gamma$.
 	The indicated step-pulse gate voltage suddenly shifts the level position at the time $t=0$.
 	(b) Dynamics after a sudden shift of the energy level from $\epsilon(t<0) = 10\Gamma$ to $\epsilon(t > 0) = -6\Gamma$ (figure adapted from Ref.~\onlinecite{Dittmann2018}).  
 	Shown are TDDFT density evolutions calculated with the non-adiabatic XC potential from Ref.~\onlinecite{Dittmann2018} (long dashed), its adiabatic counterpart (short dashed) and without XC potential (dashed-dotted lines).
 	The solid line presents a rate-equation result. Further parameters are $T = 2\Gamma$, $U=16\Gamma$.
	}
	\label{fig_setup} 
\end{figure}

As a testbed to benchmark the performance of a recently proposed non-adiabatic exchange-correlation potential \cite{Dittmann2018}, we consider a single impurity Anderson model at finite temperature, as depicted in Fig~\ref{fig_setup}(a).
The Hamiltonian is given by
\begin{align}
\label{eq_hamiltonian}
 H (t)= 
 \begin{aligned}[t]
   &\sum_\sigma \epsilon(t) d_\sigma^\dagger d_\sigma + U d_\uparrow^\dagger d_\uparrow d_\downarrow^\dagger d_\downarrow \\ &
   +\sum_{k,\sigma} \epsilon_{ k} c_{ k\sigma}^\dagger c_{ k\sigma} \\ &
   + \sum_{k,\sigma}\left( \gamma c_{k\sigma} d_{\sigma}^\dagger +  \mathrm{h.\,c.}\right),
 \end{aligned}
\end{align}
where the first, second, and third term describe the isolated impurity, the reservoir, and the reservoir-impurity coupling, respectively. 
The operators $d_\sigma^{(\dagger)}$ annihilate (create) a particle with spin $\sigma$ on the impurity site while 
$c_{k\sigma}^{(\dagger)}$ denotes the annihilation (creation) operator for a particle in quasi-momentum state $k$ and spin state $\sigma$ in the reservoir. The Coulomb repulsion of the electrons on the impurity site is described by the parameter $U$ while the electrons inside the reservoir are treated as non-interacting. The reservoir itself is in thermal equilibrium described by a chemical potential $\mu$, which is set to zero, and temperature $T$. Since we are not interested in the details of the reservoir, we employ the wide band limit, i.e.\ we assume that the density of reservoir states $\nu_0$ is a constant. This leads to a frequency independent coupling $\Gamma = 2\pi |\gamma|^2 \nu_0$ between the reservoir and the single impurity site. 

We consider two different scenarios for the energy level $\epsilon(t)$: a quench protocols where the energy level of the impurity site $\epsilon(t)=\epsilon(t<0)\Theta(-t)+\epsilon(t>0)\Theta(t)$ is rapidly changed from $\epsilon(t<0)$ to $\epsilon(t>0)$ at time $t=0$ by applying a gate voltage, and a low-amplitude harmonic oscillation, $\epsilon(t)=\bar{\epsilon} + A \sin(\omega t)$, around a mean value $\bar{\epsilon}$. 

\subsection{Time-dependent density-functional theory}
\label{sec_method_TDDFT}

In TDDFT, the interacting system defined in Eq.~\eqref{eq_hamiltonian} is simulated by an auxiliary non-interacting---Kohn-Sham (KS)---system, which has the same electron density as the interacting system. The identical densities are achieved  thanks to the Hartree (H)  and the exchange-correlation (XC) potentials in the KS system, which model electrostatic and all further interaction effects.
Along the lines of Ref.~\onlinecite{Dittmann2018}, we define the KS Hamiltonian for the interacting single-impurity Anderson model of Eq.~\eqref{eq_hamiltonian} by
\begin{align}
\label{eq_hamiltonian_KS}
 H_\mathrm{KS} (t)= 
 \begin{aligned}[t]
   &\sum_\sigma \left[\epsilon(t)+\epsilon_\mathrm{HXC}[n](t)\right] d_\sigma^\dagger d_\sigma \\ &
   +\sum_{k,\sigma} \epsilon_{ k} c_{ k\sigma}^\dagger c_{ k\sigma}
   + \sum_{k,\sigma}\left( \gamma c_{k\sigma} d_{\sigma}^\dagger +  \mathrm{h.\,c.}\right),
 \end{aligned}
\end{align}
where $\epsilon_\mathrm{HXC}[n](t)$ denotes the Hartree and XC contributions.
We assume this HXC potential to be a functional of the electron density $n$ on the impurity site.
Note that a similar modeling has been used in previous TDDFT studies of the Anderson impurity model, e.\,g., focusing on Coulomb blockade\cite{Kurth10,Evers11}, strong correlation\cite{Stefanucci11}, attractive interaction\cite{Perfetto12} as well as XC corrections to an applied bias voltage in a two-reservoir setup\cite{Kurth13,Liu15}. In all these works an adiabatic functional was employed for the HXC contribution.

\begin{figure}[t]
\begin{center}
\includegraphics[width=0.7\columnwidth]{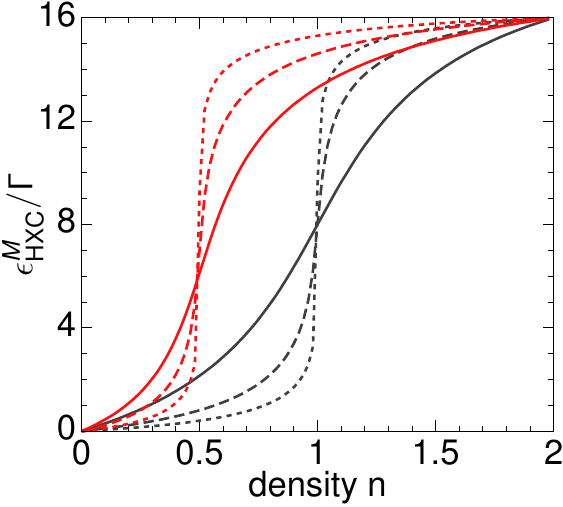}
\end{center}
	\caption{	
 	Non-adiabatic Hartree-exchange-correlation (HXC) potential $\epsilon^M_\mathrm{HXC}(n,\dot n)$ from Ref.~\onlinecite{Dittmann2018}, for $\dot n/\Gamma=0, 1$ (black and red) and 
 	temperatures $T/\Gamma=4, 2, 1$ (solid, long dashed and short dashed lines).
 	Red lines show the dynamical step appearing for $\dot n \neq 0$.
	}
	\label{fig_HXCpotential} 
\end{figure}

Ref.~\onlinecite{Dittmann2018} derives a non-adiabatic approximation for the HXC potential for the single-impurity Anderson model by using a reverse-engineering procedure based on perturbation theory in the tunnel coupling.
The result, which we write as $\epsilon_\mathrm{HXC}^M\bigl(n(t),\dot n(t),t\bigl)$, turns out to only depend on the electron density on the impurity site and its first time derivative at time $t$ and reads 
\begin{widetext}
\begin{subequations}
\label{eq_caseC_HXC}
\begin{align}
\epsilon_{\mathrm{HXC}}^{\mathrm{M}}\big(n(t),\dot{n}(t)\big)(t) &= T \log \Big\{C\big(n(t),\dot{n}(t)\big)\Big\},
\end{align}
with
\begin{align}
C(n,\dot{n}) &= \frac{2 e^{U/T}(\dot{n}+\Gamma(n-2))}
{\dot{n}+e^{U/T}\left(\dot{n}+2\Gamma (n-1)\right) 
-\sqrt{\left(\dot{n}+e^{U/T}(\dot{n}+2\Gamma(n-1))\right)^2-4e^{U/T}\left((\dot{n}+\Gamma n)^2-2\Gamma (\dot{n}+\Gamma n)\right)}} .
\end{align}
\end{subequations}
\end{widetext}
A key property of this HXC potential is a smeared-out step at half-filling, which becomes a dynamical step for nonzero $\dot n(t)$. Both is demonstrated in Fig.~\ref{fig_HXCpotential} for different temperatures, where a lower temperature leads to a sharper potential step.\cite{Stefanucci11,Evers11}
Note that the underlying perturbative expansion\cite{Schoeller96} is justified in the weak coupling/high temperature regime, where $\Gamma/T \ll 1$.
In this work we show that---as long as this constraint is satisfied---TDDFT with the non-adiabatic HXC potential $\epsilon_\mathrm{HXC}^M$ results in a highly accurate description of the electron dynamics of the single-impurity Anderson model.

In our so-called ensemble TDDFT simulations we propagate the state of the KS system in sufficiently small time steps, 
and we include a large but finite number of states for the reservoir, see Ref.~\onlinecite{Dittmann2018} for technical details.
Initially, we begin with a KS system in thermal equilibrium, and we take into account a time-dependent energy level, which is driven either by a step-pulse (Sec.~\ref{sec_results_transients}) or a low-amplitude harmonic drive 
(Sec.~\ref{sec_results_RC}), switched on at $t=0$.
During the time propagation, the value of $\epsilon_\mathrm{HXC}^M$ in the KS Hamiltonian changes continuously, based on the evolving electron density.
Notably, since the KS system is non-interacting, the numerical cost of these calculations scales linearly in simulation time, which provides an (in general) exponential speed-up compared to numerically exact methods such as 
e.\,g., the density matrix renormalization group calculations which we introduce in Sec.~\ref{sec_method_DMRG}.

Note that, although it is straightforward from a technical point of view to address more complicated quantum dot geometries or couplings then the ones considered in this paper, one has to keep in mind that the exchange correlation potential of Ref.~\onlinecite{Dittmann2018} was derived using the single-impurity Anderson model. More complicated geometries might call for a modification of this derivation, e.\,g.\ to include energy-dependent couplings between the quantum dot and the reservoir.

\subsection{Time-dependent density-matrix renormalization group}
\label{sec_method_DMRG}

In many situations, including non-equilibrium setups, interacting one-dimensional systems can be described efficiently using DMRG techniques.~\cite{White1992,Schollwock2011a} The model described in Eq.~\eqref{eq_hamiltonian} can be mapped onto a one-dimensional linear chain, if we choose to re-write the structureless reservoir as a nearest-neighbor hopping tight binding chain, described by hopping amplitude $t_h$. The resulting reservoir dispersion relation $\epsilon_k=-2t_h\cos(k)$ does not render a structureless reservoir density of states in general, which is integral to compare to the TDDFT results directly. 
However, if we choose $t_h$ large enough, such that the bandwidth $D=4t_h$ is much larger than any other energy scale of the system, then the system is effectively in the wide-band limit characterized solely by a frequency independent hybridization $\Gamma=2|\gamma|^2/t_h$. Unfortunately, the large $t_h$ limit is numerically undesirable, because as a consequence of the strong hopping, excitations propagate through the reservoir quickly. As our DMRG (in this non-translation invariant case) is set up for finite systems, information on the finiteness of the system  is imprinted on the dynamics of the impurity on increasingly smaller time-scales as $t_h$ is increased. A balance must be struck such that the results are converged with respect to both the wide-band limit as well as system size.\cite{Nghiem16} We checked this convergence by comparing results for different ratios of $t_h/\gamma$ and reservoir sizes $N$. We found that choosing $t_h/\gamma=0.05$ and $N=200$ yields converged results and thus used these values for all the DMRG calculations presented here.  

To simulate the quench dynamics we employ a two-step procedure. First, we prepare the finite temperature equilibrium state of Eq.~\eqref{eq_hamiltonian} for $H(t<0)$  with $\mu=0$ and a given temperature $T$ using the technique of purification (see Sect.~7 of Ref.~\onlinecite{Schollwock2011a}). We then employ a real time evolution algorithm propagating the system with respect to the changed Hamiltonian $H(t>0)$.\cite{Kennes2016a} In this work we use a fourth-order Suzuki-Trotter decomposition with $t_h\Delta t =0.02 $ which is small enough to give converged results. The numerical cost of this method scales in an  exponential fashion with the entanglement in the system. The control parameter encoding the entanglement growth (and hence the numerical cost) is the so-called bond-dimension. In our simulations the bond dimension is dynamically increased during the real time evolution such that we obtain numerically exact result. During the simulation the bond dimension generically grows exponentially with simulation time. As a consequence the calculation can only be carried out until the exponential growth exhaust the numerical resources available and no further progress can be made.

\section{Results}\label{sec:results}
\subsection{Transient dynamics with TDDFT and td-DMRG}
\label{sec_results_transients}

\begin{figure}[t]
\begin{center}
\includegraphics[width=\columnwidth]{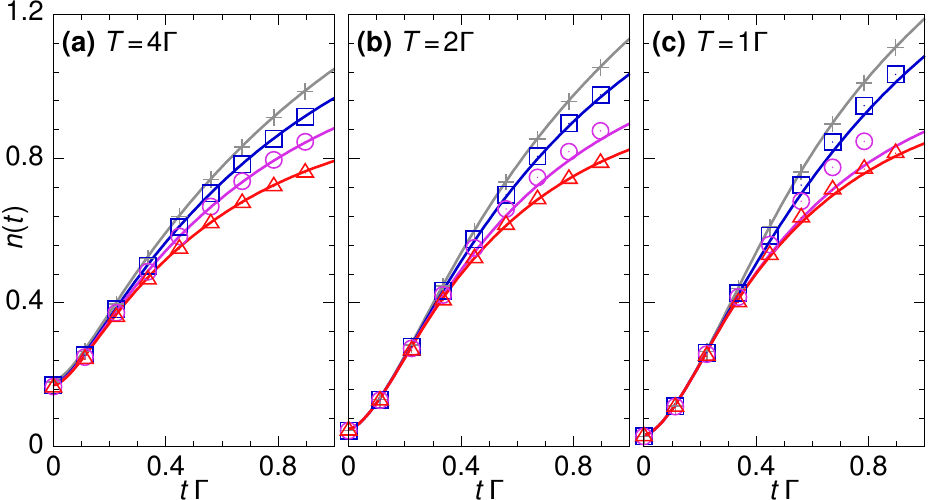}
\end{center}
	\caption{	
 	Comparison of TDDFT densities calculated using $\epsilon_{\mathrm{HXC}}^M$ (lines) with td-DMRG data (symbols).
 	The quantum dot is subject to a quench of the dot's  energy level with \mbox{$\epsilon(t<0)=10\Gamma$} and \mbox{$\epsilon(t>0)=-6\Gamma$}.
 	Top to bottom lines/symbols show different interaction strengths, with \mbox{$U/\Gamma=0,4,8,16$}, and panels (a)-(c) different temperatures.
	}
	\label{fig_tddft_dmrg_quench1} 
\end{figure}

In this section we present the results of a systematic benchmark of the non-adiabatic exchange-correlation potential put forward in Ref.~\onlinecite{Dittmann2018} by comparing to numerically exact DMRG results. 

Fig.~\ref{fig_tddft_dmrg_quench1} summarizes results for a quench of the energy level of the quantum dot from a large positive to a negative value by setting $\epsilon(t<0)=10\Gamma$ and $\epsilon(t>0)=-6\Gamma$. 
From left to right the results of the different panels show results for decreasing reservoir temperatures, $T=4\Gamma$, $T=2\Gamma$, and $T=1\Gamma$.  
Different curves in the same panel indicate different interaction strengths $U/\Gamma$. The TDDFT results are indicated by solid lines, while the DMRG results are shown as symbols. 
Increasing the interaction from $U/\Gamma=0$ (grey line and symbols) to finite and strong values the effects of interactions are clearly discernible for all curves shown. The filling of the initially almost empty quantum dot is hampered by the interactions on time-scales which are not small compared to $1/\Gamma$. The TDDFT results agree very well with the numerically exact results obtained using DMRG upto temperatures as low as $T=2\Gamma$. 
For even lower temperatures quantitative deviations are found at times large enough such that the interactions influence the quantum dot's dynamics. The discrepancy between TDDFT and DMRG is largest when the interaction strength is such that the KS energy level after the quench is close to the Fermi energy of the reservoir. We note that the qualitative behavior is in agreement for all interaction strengths and temperatures.
\begin{figure}[t]
\begin{center}
\includegraphics[width=\columnwidth]{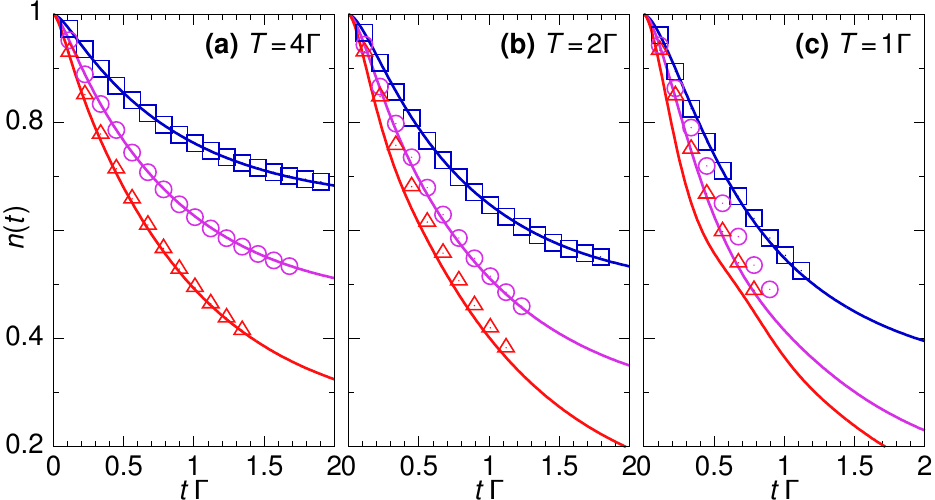}
\end{center}
	\caption{	
 	Comparison of TDDFT densities calculated using $\epsilon_{\mathrm{HXC}}^{M}$ (lines) with td-DMRG data (symbols)
 	for  a quench of the dot's  energy level with \mbox{$\epsilon(t<0)=-U/2$} and \mbox{$\epsilon(t>0)=U/2$}.
 	Top to bottom lines/symbols show different interaction strengths, with \mbox{$U/\Gamma=4,8,16$}, and panels (a)-(c) different temperatures.
	}
	\label{fig_tddft_dmrg_quench2} 
\end{figure}

As a second example we consider a quench protocol starting from the half-filled quantum dot and quenching up the energy level, such that electrons are dumped into the reservoir. The initially half-filled dot is prepared by choosing $\epsilon(t<0)=-U/2$ and then the quantum dot is quenched to $\epsilon(t<0)=U/2$. Since the strength  of the quench is measured in units of $U$ the non-interacting curve corresponds to performing no quench at all and is consequently omitted in the following. From the comparison of the TDDFT to the DMRG results shown in Fig.\ \ref{fig_tddft_dmrg_quench2} we conclude a similar picture as found above. As long as the interaction is not small and the times are large enough such that interactions actually reflect in the dynamics the agreement between the two methods is very convincing as long as the temperature is high enough. 
Here, we find again that the agreement starts to deteriorate as we approach $T=2\Gamma$ at least for the strongly interacting cases. Again the deviations are of quantitative not qualitative nature. 

From our results we summarize that the non-adiabatic exchange-correlation potential put forward in Ref.~\onlinecite{Dittmann2018} can be very useful to describe the transient dynamics of the single impurity Anderson model, 
in contrast to TDDFT approaches using adiabatic exchange-correlation potentials\cite{Stefanucci11,Stefanucci17} which fail at this task.
The regime of quantitative validity, as expected, is restricted to temperature regions which are of the order of $\Gamma$
which prohibits access to the strongly correlated Kondo-regime of pure spin-fluctuations, but nevertheless leaves a large parameter space, where this functional can be applied. 
The numerical efficiency of the TDDFT calculations ---computational cost scales only linearly instead of exponentially in simulation time--- allows to access systems and times scales beyond state-of-the-art DMRG simulation 
including intriguing questions about more complex multi-dot configurations.~\cite{Dittmann2018}

Additionally, we point out the success of steady-state density-functional theory (i-DFT) \cite{Stefanucci15,Kurth16,Stefanucci17} to describe strong correlation in the single impurity Anderson model,
which shows that DFT with an accurate exchange-correlation potential can be applied even in the Kondo regime.
We anticipate that an exchange-correlation potential which combines properties of both, the potential of Ref.~\onlinecite{Stefanucci17} and the non-adiabatic potential of Ref.~\onlinecite{Dittmann2018}, 
opens up a path towards TDDFT simulations of the dynamics in the Anderson impurity model at low temperatures.\cite{Kurth2018Jun}

\subsection{RC times with TDDFT and PT}
\label{sec_results_RC}

TDDFT also provides the proper framework to extract linear-response observables from an equilibrium DFT calculation.
The exchange-correlation potential thereby enters in terms of the exchange-correlation kernel.
This kernel is calculated from the exchange-correlation potential by a functional derivative with respect to the density, which is evaluated at the equilibrium density.
In this section, we present a second benchmark of the non-adiabatic exchange-correlation potential derived in Ref.~\onlinecite{Dittmann2018}, focusing on the linear-response dynamics of our system.
We compare the derived TDDFT data with perturbation theory in the tunnel coupling in first as well as in second order, see, e.\,g., Ref.~\onlinecite{Koenig96a,Koenig96b,Splettstoesser06,Cavaliere09,Splettstoesser10}.
This comparison highlights the conceptual difference between a perturbation-theory description on one hand, and, on the other hand, TDDFT with a HXC potential obtained in perturbation theory.

We investigate a small-amplitude harmonic drive of the energy level $\epsilon(t)=\bar{\epsilon} + A \sin(\omega t)$, around a mean value $\bar \epsilon$, and we calculate the finite-frequency admittance $G(\omega) = \left.\frac{\partial I(\omega)}{\partial \epsilon (\omega)}\right|_\mathrm{eq}$.
As pointed out by B\"uttiker et al.~\cite{Buettiker93}, it is instructive to compare the low-frequency part of this admittance to the admittance of an RC circuit, which is the classical analog of the system sketched in Fig.~\ref{fig_setup}~(a).
The low-frequency expansion, $G(\omega) = -i\omega C + \omega^2 C^2 R$, defines a charge-relaxation resistance $R$ and an electrochemical capacitance $C$ for the single-impurity Anderson model.
In our benchmark, we compare RC times, $\tau_\mathrm{RC}=RC$, obtained in TDDFT with results from second-order perturbation theory in the tunnel-coupling strength.\cite{Splettstoesser10}
Applying TDDFT linear-response theory, we find that $R$ and $C$ are connected to the values $R_\mathrm{KS}$ and $C_\mathrm{KS}$ of the auxiliary Kohn-Sham system by
\begin{align}
\label{eq_R_Dyson}
R &= \frac{R_\mathrm{KS}+f_\mathrm{HXC}^{(1)}\bigl(n_\mathrm{eq}\bigl)}{\left[1+C_\mathrm{KS}f_\mathrm{HXC}^{(0)}\bigl(n_\mathrm{eq}\bigl)\right]^2}\,,\\
\label{eq_C_Dyson}
C &= \frac{C_\mathrm{KS}}{1+C_\mathrm{KS}f_\mathrm{HXC}^{(0)}}\,.
\end{align}
In these equations, the Hartree-exchange-correlation kernel derived from the non-adiabatic exchange-correlation potential of Ref.~\onlinecite{Dittmann2018} is written as~\cite{Dittmann18P}
\begin{align}
\begin{aligned}
 \label{eq_fhxc}
 f_\mathrm{HXC}^{M}\bigl(n_\mathrm{eq},\omega\bigl) &= f_\mathrm{HXC}^{(0)}\bigl(n_\mathrm{eq}\bigl) - i \omega f_\mathrm{HXC}^{(1)}\bigl(n_\mathrm{eq}\bigl),
\end{aligned}
\end{align}
with $n_\mathrm{eq}$ being the equilibrium density on the impurity site, and where we abbreviate
\mbox{$f_\mathrm{HXC}^{(0)}\bigl(n_\mathrm{eq}\bigl) = \frac{\partial \epsilon^{M}_\mathrm{HXC}(n,\dot{n})}{\partial n}\biggl|_\mathrm{eq}$} and
\mbox{$f_\mathrm{HXC}^{(1)}\bigl(n_\mathrm{eq}\bigl) = \frac{\partial \epsilon^{M}_\mathrm{HXC}(n,\dot{n})}{\partial \dot{n}}\biggl|_\mathrm{eq}$}.
The equilibrium density is obtained by self-consistently solving the density expression \mbox{$n_\mathrm{eq} = \frac{\Gamma}{\pi} \int_{-\infty}^{\infty} \frac{f(E)}{\left(\epsilon_\mathrm{KS}[n_\mathrm{eq}]-E\right)^2+\Gamma^2/4}dE$},
with the Fermi function \mbox{$f(E)=1/(1+e^{E/T})$}. 
Note that the finite-frequency admittance $G_\mathrm{KS}(\omega)$ and thus $R_\mathrm{KS}$ and $C_\mathrm{KS}$ are calculated exactly for the non-interacting Kohn-Sham system.\cite{Jauho94}

\begin{figure}[t]
\begin{center}
\includegraphics[width=\columnwidth]{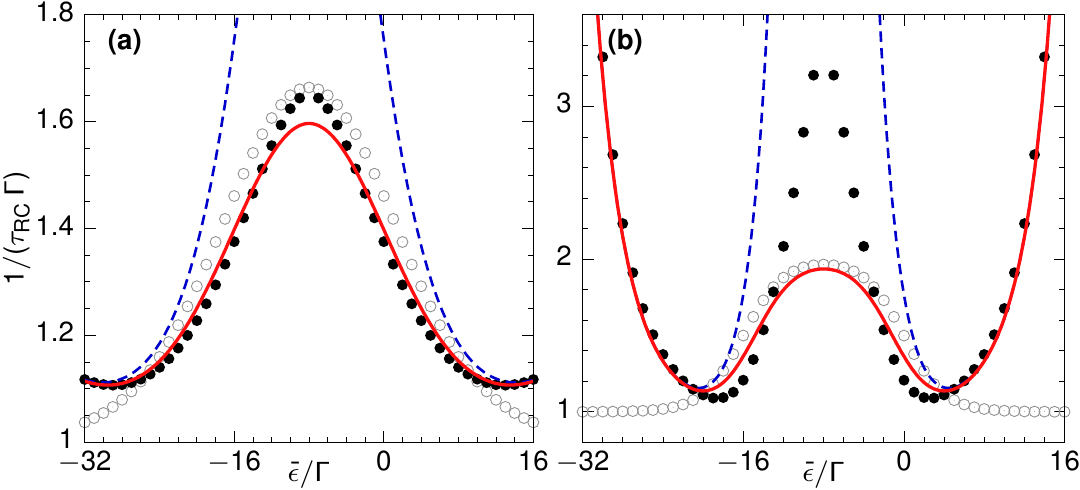}
\end{center}\vspace{-5mm}
	\caption{Inverse $RC$ times for two temperatures (a) \mbox{$T=5\Gamma$} and (b) \mbox{$T=2 \Gamma$}.
	Shown are the $RC$ times calculated in TDDFT using the HXC kernels $f_\mathrm{HXC}^{M}$ (solid lines) and \mbox{$f_\mathrm{HXC}^{\mathrm{ad}}=f_\mathrm{HXC}^M|_{\omega=0}$} (dashed lines),
	as well as the $RC$ times derived in first-order and second-order perturbation theory in the tunnel-coupling (empty and filled symbols), see Ref.~\onlinecite{Splettstoesser10}.
	The interaction strength is set to \mbox{$U = 16 \Gamma$}.
	}
	\label{fig_RCtimes}  
\end{figure}

In Fig.~\ref{fig_RCtimes} we present $RC$ times calculated in TDDFT (lines) for two different temperatures as indicated.
Let us first focus on panel (a), where a higher temperature is applied ($T=5\Gamma$).
The TDDFT result based on $f_\mathrm{HXC}^M$ is plotted as the solid line, and it is compared to first-order as well as second-order perturbation-theory results (empty and filled symbols).\cite{Splettstoesser10}
We observe that the first-order result shows strong deviations compared to the TDDFT data, while the second-order perturbative data agrees much better with the TDDFT result over a large range of working points $\bar \epsilon$.
This is particularly interesting, considering that the underlying HXC potential, $\epsilon_\mathrm{HXC}^M$, is derived using a first-order perturbative expansion in the reservoir-dot coupling.
The reason why the TDDFT data, nevertheless, can match second-order perturbation theory results, is that no perturbation-theory expansion is employed to solve the KS auxiliary system.
Only interaction effects are modeled with an HXC potential motivated from a first-order perturbation theory, while the time evolution in the non-interacting KS system includes all orders in the tunnel coupling.
Although it is beyond the present work to develop an HXC potential based on higher-order perturbation theory and check whether the agreement further improves compared to higher-order perturbation theory results, we believe that deviations between TDDFT and second-order perturbation theory 
stem from the fact that interaction plays a more dominant role in the dynamics.
In Fig.~\ref{fig_RCtimes}~(a), we find that this is the case for working points close to electron-hole symmetry, \mbox{$\bar \epsilon \approx -\frac{U}{2}$}.
As a further comparison, we also show RC times calculated with the adiabatic HXC potential which is related to $\epsilon_\mathrm{HXC}^M$ (and which is obtained by setting $\dot n$ to zero in the expression for $\epsilon_\mathrm{HXC}^M$).
This additional data is shown as the dashed line.
We find that  when the impurity is close to zero or double occupation, \mbox{$\bar \epsilon \gg 0$} or \mbox{$\bar\epsilon \ll -U$}, 
the adiabatic description suffices, since the interaction plays a subdominant role in these regions.
In contrast, the RC times are strongly overestimated when the system reaches single occupation.

A similar calculation performed at a lower temperature, $T=2\Gamma$, is presented in Fig.~\ref{fig_RCtimes}~(b).
Here, the impact of second-order tunneling on the dynamics is more pronounced.
In panel~(b), the limitation of $\epsilon_\mathrm{HXC}^M$ is evident: the potential leads to an over-fitting of the TDDFT data (solid line) with the first-order perturbation-theory result (empty symbols), 
in particular close to the electron-hole symmetric point, see also the center of panel (a).
This result agrees with our findings in Sec.~\ref{sec_results_transients}, namely that $\epsilon_\mathrm{HXC}^M$ is limited to TDDFT calculations at high temperatures and weak coupling, $\Gamma/T \ll 1$.
To also reach lower temperatures, further research is necessary in order to find modifications of this HXC potential which correctly account for interaction effects in this regime.

\section{Conclusion}\label{sec:conclusions}
We performed a comparative study in which we benchmarked the validity range of the TDDFT approach based on a non-adiabatic Hartree-exchange-correlation potential put forward in Ref.~\onlinecite{Dittmann2018} by comparing to (a) a numerically exact DMRG based approach as well as (b) results obtained approximately from second-order perturbation theory. In both cases, we found that the  accuracy of the non-adiabatic functional is better than expected from the derivation of the functional.

By comparing to DMRG we have found that the dynamics of a single-impurity Anderson model is accurately described by the proposed non-adiabatic TDDFT approach at least in the regime of sufficiently high temperature. Therefore, it provides access to the dynamics in this parameter regime by a method that is (at least in general cases) exponentially faster than competitive numerically exact approaches. In other words, the TDDFT calculation does not suffer from the Markov approximation which was made in the derivation of the approximate functional due to the time-propagation of the Kohn-Sham system being numerically exact. However, the very low temperature regime of strongly correlated Kondo physics seems to be off-limits to this specific approximation. 

We have then compared to results obtained with perturbation theory upto second order in the tunnel coupling. We find a significant increase in agreement between perturbation theory and TDDFT as the second order is included compared to the first order results. This shows that our TDDFT approach includes important processes of second order although the exchange-correlation potential used was motivated from first order. Again, the added accuracy can be ascribed to the exact time-propagation of the Kohn-Sham system.

The systematic benchmark of non-adiabatic exchange correlation potentials in TDDFT calculations is the first step along the route to many pressing issues. The reservoir coupled to the single-impurity model imprints relaxation onto the dot degrees of freedom. Including incoherent relaxation processes from scattering in conventional TDDFT simulations proves difficult so far. One way to proxy this relaxation process could be via explicit (and phenomenological) coupling to particle reservoirs as described in this work. Furthermore, impurity problems as studied here have gained tremendous attention in condensed matter physics, also because of the role they play in dynamical mean field studies. Within the DMFT approach a (e.g.\ time-dependent) solution of an impurity problem is required as an input. So far methods which are able to tackle complicated multi-orbital impurities and/or spin-orbit coupling are scarce, but urgently needed. Future studies should thus address the issue of multi-orbital impurity problems (or other situations) from a TDDFT perspective using non-adiabatic exchange correlation potentials.


\textit{Acknowledgements.} 
We acknowledge funding from the DFG via RTG1995 (N.D., N.H.) and the Deutsche Forschungsgemeinschaft through the Emmy Noether program (KA 3360/2-1) (D.M.K.).
Simulations were performed with computing resources granted by RWTH Aachen University under projects rwth0013 and prep0010.
\bibliographystyle{apsrev4-1}
\bibliography{cite}
\end{document}